%Paper: solv-int/9412002
%From: Dr "P.A" Clarkson <clarkson@maths.ex.ac.uk>
%Date: Fri, 9 Dec 94 11:48:03 GMT

\magnification=\magstephalf
\nopagenumbers
\hsize=6.5 truein\vsize=9.5 truein
\hfuzz=2pt\vfuzz=4pt
\pretolerance=5000
\tolerance=5000
\parskip=0pt plus 1pt
\parindent=16pt
\font\fourteenrm=cmr10 scaled \magstep2
\font\fourteeni=cmmi10 scaled \magstep2
\font\fourteenbf=cmbx10 scaled \magstep2
\font\fourteenit=cmti10 scaled \magstep2
\font\fourteensy=cmsy10 scaled \magstep2
\font\bbf=cmbx10 scaled \magstep1
\font\big=cmr10 scaled \magstep1

\font\sixrm=cmr6

\font\eightrm=cmr8
\font\eighti=cmmi8
\font\eightbf=cmbx8
\font\eightit=cmti8

\font\eightsy=cmsy8
\font\sixrm=cmr6
\font\sixi=cmmi6
\font\sixsy=cmsy6

\def\tenpoint{\def\rm{\fam0\tenrm}%
 \textfont0=\tenrm \scriptfont0=\sevenrm
	   \scriptscriptfont0=\fiverm
 \textfont1=\teni \scriptfont1=\seveni
	   \scriptscriptfont1=\fivei
 \textfont2=\tensy \scriptfont2=\sevensy
	   \scriptscriptfont2=\fivesy
 \textfont3=\tenex  \scriptfont3=\tenex
	   \scriptscriptfont3=\tenex
 \textfont\itfam=\tenit \def\it{\fam\itfam\tenit}%
 \textfont\slfam=\tensl \def\sl{\fam\slfam\tensl}%
 \textfont\bffam=\tenbf \scriptfont\bffam=\sevenbf
	      \scriptscriptfont\bffam=\fivebf
	      \def\bf{\fam\bffam\tenbf}%
 \normalbaselineskip=20 truept
 \setbox\strutbox=\hbox{\vrule height14pt depth6pt width0pt}%
 \let\sc=\eightrm \normalbaselines\rm}
\def\eightpoint{\def\rm{\fam0\eightrm}%
 \textfont0=\eightrm \scriptfont0=\sixrm
	   \scriptscriptfont0=\fiverm
 \textfont1=\eighti \scriptfont1=\sixi
	   \scriptscriptfont1=\fivei
 \textfont2=\eightsy \scriptfont2=\sixsy
	   \scriptscriptfont2=\fivesy
 \textfont3=\tenex  \scriptfont3=\tenex
	   \scriptscriptfont3=\tenex
 \textfont\itfam=\eightit \def\it{\fam\itfam\eightit}%
 \textfont\bffam=\eightbf \def\bf{\fam\bffam\eightbf}%
 \normalbaselineskip=16 truept
 \setbox\strutbox=\hbox{\vrule height11pt depth5pt width0pt}}
\def\fourteenpoint{\def\rm{\fam0\fourteenrm}%
 \textfont0=\fourteenrm \scriptfont0=\tenrm
	   \scriptscriptfont0=\eightrm
 \textfont1=\fourteeni \scriptfont1=\teni
	   \scriptscriptfont1=\eighti
 \textfont2=\fourteensy \scriptfont2=\tensy
	   \scriptscriptfont2=\eightsy
 \textfont3=\tenex  \scriptfont3=\tenex
	   \scriptscriptfont3=\tenex
 \textfont\itfam=\fourteenit \def\it{\fam\itfam\fourteenit}%
 \textfont\bffam=\fourteenbf \scriptfont\bffam=\tenbf
	       \scriptscriptfont\bffam=\eightbf
	       \def\bf{\fam\bffam\fourteenbf}%
 \normalbaselineskip=24 truept
 \setbox\strutbox=\hbox{\vrule height17pt depth7pt width0pt}%
 \let\sc=\tenrm \normalbaselines\rm}

\def\today{\number\day\ \ifcase\month\or
 January\or February\or March\or April\or May\or June\or
 July\or August\or September\or October\or November\or December\fi
 \space \number\year}
\newcount\secno   %section number
\newcount\subno   %number of subsection
\newcount\subsubno  %number of subsubsection
\newcount\appno   %appendix number
\newcount\tableno  %table number
\newcount\figureno  %figure number
\normalbaselineskip=15 truept
\baselineskip=15 truept

\def\title#1#2
  {\vglue0.5truein
  {\centerline{\bbf #1}\smallskip\centerline{\bbf #2}
  \vskip 0.5truein}}
\def\author#1{\centerline{\big #1}\medskip}
\def\address#1#2{\centerline{\sc #1}\smallskip
\centerline{{\it Email address}: {\tt #2}}\medskip}

\def\shorttitle#1
  {\vfill
  \noindent \rm Short title: {\sl #1}\par
  \medskip}
\def\pacs#1
  {\noindent \rm PACS number(s): #1\par
  \medskip}
\def\jnl#1
  {\noindent \rm Submitted to: {\sl #1}\par
  \medskip}
\def\date
  {\centerline{Date: \today}\medskip}
\def\keyword#1
  {\bigskip
  \noindent {\bf Keyword abstract: }\rm#1}
\def\abstract#1{\vskip 0.5truein
  {\baselineskip=12pt\narrower\noindent{\sc Abstract}. \rm#1}}

\def\appendix#1
  {\vskip0pt plus.1\vsize\penalty-250
  \vskip0pt plus-.1\vsize\vskip24pt plus12pt minus6pt
  \subno=0 \eqnno=0
  \global\advance\appno by 1
  \noindent {\bf Appendix \the\appno. #1\par}
  \bigskip
  \noindent}
\def\subappendix#1
  {\vskip-\lastskip
  \vskip36pt plus12pt minus12pt
  \bigbreak
  \global\advance\subno by 1
  \noindent {\sl \the\appno.\the\subno. #1\par}
  \nobreak
  \medskip
  \noindent}
\def\ack
  {\vskip0pt plus.1\vsize\penalty-250
  \vskip0pt plus-.1\vsize\vskip18pt plus9pt minus6pt
  \bigbreak
  \noindent{\bf Acknowledgements\par}
  \nobreak
  \bigskip
  \noindent}

\def\tabcaption#1
  {\global\advance\tableno by 1
  \noindent {\bf Table \the\tableno.} \rm#1\par
  \bigskip}

\def\figcaption#1
  {\global\advance\figureno by 1
  \noindent {\bf Figure \the\figureno.} \rm#1\par
  \bigskip}
\def\references
   {\vskip0pt plus.1\vsize\penalty-250
  \vskip0pt plus-.1\vsize\vskip18pt plus9pt minus6pt
   %\vskip .5in \vfill\eject
   \noindent{\bf References}\par
   \parindent=0pt
   \bigskip}

\def\frac#1#2{{\displaystyle{#1 \over #2}}}

\def\d{{\rm d}}

\def\i{\ifmmode{\rm i}\else\char"10\fi}
\def\case#1#2{{\textstyle{#1\over #2}}}

\def\etal{{\sl et al.\/}\ }
\catcode`\@=11
\def\ind{\hbox to 5pc{}}
\def\eq(#1){\hfill\llap{(#1)}}

\def\deqn#1{\displ@y\halign{\hbox to \displaywidth
  {$\@lign\displaystyle##\hfil$}\crcr #1\crcr}}
\def\indeqn#1{\displ@y\halign{\hbox to \displaywidth
  {$\ind\@lign\displaystyle##\hfil$}\crcr #1\crcr}}
\def\indalign#1{\displ@y \tabskip=0pt
 \halign to\displaywidth{\ind$\@lign\displaystyle{##}$\tabskip=0pt
  &$\@lign\displaystyle{{}##}$\hfill\tabskip=\centering
  &\llap{$\@lign##$}\tabskip=0pt\crcr
  #1\crcr}}
\catcode`\@=12

%%% End of IOP macros

\font\sc=cmcsc10

\font\nit=cmti9
\font\nrm=cmr9
\font\nbf=cmbx9
\font\nsl=cmsl9

\font\shell=msbm10

\def\N{{\hbox{\shell N}}}
\def\Na{{\hbox{\shell N}}}

\def\Z{{\hbox{\shell Z}}}

\def\tx{\textstyle}

\def\fr#1#2{{\displaystyle{#1\over#2}}}
\def\tfr#1#2{{\tx{#1\over#2}}}

\def\eps{\varepsilon}

\def\p{Pain\-lev\'e}
\def\bk{B\"ack\-lund}
\def\bt{B\"ack\-lund transformation}
\def\bts{B\"ack\-lund transformations}

\def\erfc{\mathop{\rm erfc}\nolimits}

\newcount\exno
\newcount\secno  %number of section
\newcount\subno  %number of subsection
\newcount\subsubno  %number of subsubsection
\newcount\figno
\newcount\tableno

\def\section#1
  {\vskip0pt plus.1\vsize\penalty-250
  \vskip0pt plus-.1\vsize\vskip18pt plus9pt minus6pt
  \subno=0 \exno=0 \figno=0 \eqnno=0
  \global\advance\secno by 1
  \centerline{\bf\the\secno. #1}
  \medskip}
\def\subsection#1
  {\vskip-\lastskip
  \exno=0 \caseno=0 \tableno=0 \subsubno=0
  \vskip15pt plus4pt minus4pt
  \bigbreak
  \global\advance\subno by 1
  {\bf \the\secno.\the\subno. #1.}\enskip}

\def\subsubsection#1
  {\vskip-\lastskip
  \exno=0 \caseno=0
   \vskip4pt plus2pt minus2pt
  \bigbreak \global\advance\subsubno by 1
  {\sl \the\secno.\the\subno.\the\subsubno\enskip #1. }}

\def\=#1{{\bf\bar{\mit#1}}}
\def\^#1{{\widehat{\mit#1}}}
\def\~#1{{\widetilde{\mit#1}}}
\def\He{\hbox{He}}

\newcount\eqnno
\def\sen{\the\secno}%{\the\secno.\the\subno}%
\def\eqn#1{\global\advance\eqnno by 1
      \eqno(\the\secno.\the\eqnno)
      \expandafter \xdef\csname #1\endcsname
      {\the\secno.\the\eqnno}\relax }
\def\eqnn#1{\global\advance\eqnno by 1
      (\the\secno.\the\eqnno)
      \expandafter \xdef\csname #1\endcsname
      {\the\secno.\the\eqnno}\relax }
\def\eqnm#1#2{\global\advance\eqnno by 1
      (\the\secno.\the\eqnno\hbox{#2})
      \expandafter \xdef\csname #1\endcsname
      {\the\secno.\the\eqnno}\relax }
\def\eqnr#1{(\the\secno.\the\eqnno\hbox{#1})}

\newcount\caseno
\def\case#1{\vskip-\lastskip
  \vskip4pt plus2pt minus2pt
  \bigbreak \global\advance\caseno by 1
\noindent\underbar{{\sc Case}
\the\secno.\the\subno.\the\caseno}\enskip{#1}. }

\def\Case#1{\vskip-\lastskip
  \vskip4pt plus2pt minus2pt
  \bigbreak \global\advance\caseno by 1
\noindent {\sc
Case}\ \the\secno.\the\subno.\the\exno{\romannumeral\the\caseno}
\enskip \underbar{#1}.\quad}

\def\p{Pain\-lev\'e}
\def\Pi{first \p\ equation}

\font\sevensy=cmsy7
\font\sevenit=cmti7
\font\sevenrm=cmr7
\font\sevenbf=cmbx7
\def\sevenpoint{\def\rm{\fam0\sevenrm}
\textfont0=\sevenrm \textfont1=\sevenit \textfont2=\sevensy}
\def\sdag{{\sevenpoint\dag}}
\def\sddag{{\sevenpoint\ddag}}

\def\cite#1{[{\bf#1}]}
\def\sqb{\sqrt{-2 \beta}}
\def\widehat{\mathaccent"362}
\def\widetilde{\mathaccent"365}
\def\~#1{{\bf\widetilde{\mit#1}}}
\def\^#1{{\bf\widehat{\mit#1}}}

\def\hide#1{}

\headline={\ifnum\pageno>1
\ifodd\pageno\rightheadline\else\leftheadline\fi
\else\hfil\fi}
\def\rightheadline{\tenrm\hfil {\sc peter a.\ clarkson and andrew p.\
bassom}\hfil\folio}
\def\leftheadline{\tenrm\hfil {\sc exact solutions of
the fourth painlev\'e equation and discrete equations}\hfil\folio}

\newcount\refno
\refno=0

\def\refdpp#1#2#3#4{\global\advance\refno by 1
\expandafter \xdef\csname #1\endcsname {\the\refno}\relax}{}{}{}
\def\refpp#1#2#3#4{\global\advance\refno by 1
\expandafter \xdef\csname #1\endcsname {\the\refno}\relax}{}{}{}{}
\def\refjl#1#2#3#4#5#6{\global\advance\refno by 1
\expandafter \xdef\csname #1\endcsname {\the\refno}\relax}{}{}{}{}{}{}
\def\refbk#1#2#3#4#5{\global\advance\refno by 1
\expandafter \xdef\csname #1\endcsname {\the\refno}\relax}{}{}{}{}
\def\refcf#1#2#3#4#5#6#7{\global\advance\refno by 1
\expandafter \xdef\csname #1\endcsname {\the\refno}\relax}{}{}{}{}{}{}

\refbk{refAC}{M.J. Ablowitz and P.A. Clarkson}{Solitons, Nonlinear
Evolution Equations and Inverse Scattering, {\frenchspacing\it
L.M.S.\ Lect.\
Notes Math.\/}, {\bf149}}{C.U.P., Cambridge}{1991}
\refjl{refASa}{M.J. Ablowitz and H. Segur}{Phys. Rev.
Lett.}{38}{1103--1106}{1977}
\refbk{refASte}{M. Abramowitz and I.A. Stegun}{Handbook of Mathematical
Functions}{Dover, New York}{1965}
\refjl{refBC}{A.P. Bassom and P.A. Clarkson}{Phys. Lett.
A}{194}{358--370}{1994}
\refjl{refBCH}{A.P. Bassom, P.A. Clarkson and A.C. Hicks}{IMA J. Appl.
Math.}{50}{167--193}{1993}
\refjl{refBCHI}{A.P. Bassom,\ P.A. Clarkson\ and A.C. Hicks}{Stud.
Appl. Math.}{}{to appear}{1995}
\refdpp{refBCHII}{A.P. Bassom,\ P.A. Clarkson\ and A.C. Hicks}
{``On the application of solutions of the fourth Painlev\'e equation to
various
physically motivated nonlinear partial differential
equations'', preprint {\bf M94/32}, Department of Mathematics,
University of
Exeter}{1994}
\refjl{refBCHM}{A.P. Bassom, P.A. Clarkson, A.C. Hicks and J.B.
McLeod}{Proc.
R. Soc. Lond. A}{437}{1--24}{1992}
\refjl{refBK}{E. Brezin and V. Kazakov}{Phys. Lett.
B}{236}{144--150}{1990}
\refjl{refPAC}{P.A. Clarkson}{Europ. J. Appl. Math.}{1}{279--300}{1990}
\refjl{refDoug}{M.R. Douglas}{Phys. Lett. B}{238}{176--180}{1990}
\refjl{refFGR}{A.S. Fokas, B. Grammaticos and A. Ramani}{J. Math.
Anal.
Appl.}{180}{342--360}{1993}
\refjl{refFIK}{A.S. Fokas, A.R. Its and A.V. Kitaev}{Commun. Math.
Phys.}{142}{313--344}{1991}
\refjl{refFMA}{A.S. Fokas, U.\ Mugan and M.J.
Ablowitz}{Physica}{30D}{247--283}{1988}
\refjl{refGNPRS}{B. Grammaticos, F.W. Nijhoff, V. Papageorgiou, A.
Ramani and J. Satsuma}{Phys. Lett. A}{185}{446--452}{1994}
\refcf{refGR}{B. Grammaticos and A. Ramani}{Kluwer, Dordrecht,
1993}{Applications of Analytic and Geometric Methods to Nonlinear
Differential
Equations}{P.A.\ Clarkson, ed.}{{NATO ASI Series C\/},
{\bf413}}{299--313}
\refjl{refRGV}{B. Grammaticos,\ A. Ramani and V. Papageorgiou}
{Phys. Rev. Lett.}{67}{1825--1828}{1991}
\refjl{refGromak}{V.I. Gromak}{Diff. Eqns.}{14}{1510--1513}{1977}
\refjl{refGM}{D.J. Gross and A.A. Migdal}{Phys. Rev.
Lett.}{64}{127--130}{1990}
\refbk{refInce}{E.L. Ince}{Ordinary Differential Equations}{Dover, New
York}{1956}
\refjl{refKOSGR}{K. Kajiwara, Y. Ohta, J. Satsuma, B. Grammaticos and
A.
Ramani}{J. Phys. A: Math. Gen.}{27}{915--922}{1994}
\refpp{refKitpc}{A.V. Kitaev}{private communication}{1991}
 \refjl{refLukash}{N.A. Lukashevich}{Diff. Eqns.}{3}{395--399}{1967}
\refjl{refMurata}{Y. Murata}{Funkcial. Ekvac.}{28}{1--32}{1985}
\refjl{refNP}{F.W. Nijhoff and V. Papageorgiou}{Phys. Lett.
A}{153}{337--344}{1991}
\refjl{refPNGR}{V. Papageorgiou, F.W. Nijhoff, B. Grammaticos and A.
Ramani}{Phys. Lett. A}{164}{57--64}{1992}
\refjl{refPS}{V. Periwal and D. Shewitz}{Phys. Rev.
Lett.}{64}{1326--1329}{1990}
\refjl{refRGH}{A. Ramani, B. Grammaticos and J. Hietarinta}{Phys. Rev.
Lett.}{67}{1829--1832}{1991}
\refbk{refSchiff}{L.I. Schiff}{Quantum Mechanics}{McGraw-Hill, New
York}{1955}
\refjl{refTGR}{K.M. Tamizhmani,\ B. Grammaticos and A. Ramani}{Lett.
Math.
Phys.}{29}{49--54}{1993}

\title{\bk\ Transformations and Hierarchies of Exact Solutions for
the}{Fourth
\p\ Equation and  their Application to Discrete Equations}
\author{Peter A.\ Clarkson and Andrew P.\ Bassom}

\abstract{In this paper we describe \bk\ transformations and
hierarchies of
exact solutions for the fourth \p\ equation (PIV)
$${\d^2 w\over\d z^2}={1\over2w}\left(\d w\over\d z\right)^2 +
{{3\over2}}w^3 + 4zw^2 + 2(z^2-\alpha)w+{\beta\over w},\eqno(1){\hbox
to
16pt{\hfill}}$$ with
$\alpha$, $\beta$ constants. Specifically, a nonlinear superposition
principle
for PIV,  hierarchies of solutions expressible in terms of
complementary error or
parabolic cylinder functions  as well as rational solutions will be
derived.
Included amongst these hierarchies are solutions of (1) for which
$\alpha=\pm\tfr12n$ and $\beta=-\tfr12n^2$, with $n$ an integer.  These
particular forms arise in quantum gravity and also satisfy a discrete
analogue
of the first \p\ equation. We also obtain a number of exact
solutions of the discrete fourth Painlev\'e equation
$$x_{n+1}x_{n-1}+x_n(x_{n+1}+x_{n-1})=
{-2z_nx_n^3+(\eta-3\delta^{-2}-z_n^2)x_n^2+\mu^2\over
(x_n+z_n+\gamma)(x_n+z_n-\gamma)},\eqno(2){\hbox to 16pt{\hfill}}$$}%
{\narrower\noindent\baselineskip=12pt where $z_n=n\delta$ and $\eta$,
$\delta$,
$\mu$ and $\gamma$ are constants,  which,  in an appropriate limit,
reduces to
PIV (1). A suitable factorisation of (2) facilitates the identification
of a
number of solutions which take the form of ratios of two polynomials in
the
variable $z_n$ and the limits of these solutions yield rational
solutions of
(1).\par }

\baselineskip=13pt

\section{Introduction}
The six \p\ equations (PI--PVI) were first derived around the turn
of the century in an investigation by \p\ and his colleagues into which
second-order ordinary differential equations have the property that the
singularities other than poles of any of the solutions are independent
of the
particular solution and so are dependent only upon the equation (cf.,
\cite{\refInce}); this property is now known as the
{\it\p\ property\/}. There
has  been considerable interest in \p\ equations over the last few
years
primarily due to the fact that they arise as reductions of soliton
equations
solvable by inverse scattering as first demonstrated by Ablowitz
\&\ Segur
\cite{\refASa}. Although  first discovered from strictly mathematical
considerations, the \p\ equations have appeared  in various of physical
applications (cf., \cite{\refAC} and the references therein). The
\p\ equations
may also be thought of as nonlinear analogues of the classical special
functions
though they are known to be transcendental since their solution is not
expressible in terms of elementary functions. However rational
solutions and
one-parameter families of solutions of the \p\ equations which can be
written
in terms of special functions are known to
exist for particular values of the parameters. For example, there exist
solutions of PII, PIII and PIV that are expressed in terms of Airy,
Bessel and
parabolic cylinder functions, respectively (cf., \cite{\refGromak}).

Recently there has been considerable interest in integrable mappings
and
discrete systems, including discrete analogues of the \p\ equations.
Some of
these mapping and discrete equations arise in physical applications.
For
example, a discrete analogue of PI (d-PI) arose in the study of the
partition
function in a two-dimensional model of quantum gravity
\cite{\refBK,\refDoug,\refFIK,\refGM}. Subsequently a discrete analogue
of PII
(d-PII) was derived in \cite{\refNP,\refPS}  and later discrete
analogues of
PIII--PV (d-PIII--d-PV) were obtained by Ramani
\etal \cite{\refRGH} using the method of singularity confinement
\cite{\refRGV}; for further details on the derivation of the discrete
\p\
equations see, for example, \cite{\refFGR,\refGR}. One important result
of the
investigations is that the form of the discrete \p\ equations is not
unique
since there exist several possible discrete analogues of the
\p\ equations.
Kajiwara \etal \cite{\refKOSGR} and Grammaticos \etal \cite{\refGNPRS}
have
derived exact solutions of d-PII and d-PIII in terms of discrete Airy
and
discrete Bessel functions, respectively, in analogue to the
aforementioned
results for the associated continuous \p\ equations. We further remark
that Lax
pairs and iso\-mono\-dromic deformation problems are known to exist for
d-PI
\cite{\refFIK}, d-PII \cite{\refNP} and d-PIII \cite{\refPNGR}.
However, at
present, there is no discrete analogue of PVI, nor are there Lax pairs
for the
versions of d-PIV and d-PV derived in \cite{\refGNPRS}.

In recent work \cite{\refBCHI,\refBCHII}, we have been concerned with
the
investigation of \bts\ and exact solutions for fourth \p\ equation
(PIV)
$$ww'' = \tfr12\left(w'\right)^2+\tfr32w^4 +
4zw^3 + 2(z^2-\alpha)w^2 +{\beta},\qquad '\equiv\d/\d z,\eqn{eqPIV}$$
where $\alpha$ and $\beta$ are arbitrary constants, together
with an examination of various applications of these solutions to
several
physically motivated nonlinear partial differential equations.  In
\cite{\refBCHI} we demonstrated how all known exact solutions of
(\eqPIV) can be
categorised into one of three families; in two of these solutions can
be
determined in terms of the complementary error and parabolic cylinder
functions
whilst the third family consists of solutions which can be expressed as
the
ratio of two polynomials in $z$.

In \S2 we review the known \bts\ for PIV (\eqPIV). In particular it is
shown
that there exist three-term recurrence relations, or {\it nonlinear
superposition formulae}, for PIV, which can be effectively used for
the derivation of solution hierarchies of PIV involving {\it
algebraic\/} manipulations alone. In \S3 we briefly discuss hierarchies
of exact
solutions for PIV including one-parameter families of solutions
expressible in
terms of parabolic cylinder functions, solutions expressible in terms
of
complementary error functions and rational solutions. In \S4 we show
that one
family of solutions for PIV, the so-called half-integer hierarchy,
generate
solutions of the discrete first \p\ equation and are relevant to
problems in
two-dimensional quantum gravity.

In \S5 we discuss some new solutions of the discrete fourth
\p\ equation (d-PIV) given by the three-point, non-autonomous mapping
$$x_{n+1}x_{n-1}+x_n(x_{n+1}+x_{n-1})=
{-2z_nx_n^3+(\eta-3\delta^{-2}-z_n^2)x_n^2+\mu^2\over
(x_n+z_n+\gamma)(x_n+z_n-\gamma)},\eqn{eqI}$$
with the variable $x_n$ to be found in terms of $z_n\equiv
n\delta+\zeta$.
This mapping is identified as the discretised version of PIV. This is
observed
by taking the limit of (\eqI) as
$\delta\to 0$, with $\gamma=1/\delta$ and $\eta$ and $\mu$ finite. This
process yields (\eqPIV) with the parameters $\alpha$ and $\beta$ in
that
equation related to $\eta$ and $\mu$ according to
$\alpha=\tfr14\eta$ and $\beta=-\tfr12\mu^2$.
Since d-PIV reduces to (\eqPIV) in the appropriate limit then it can be
expected that exact solutions
of (\eqI) exist which should tend to known continuous solutions
in the same limit. Exact solutions of d-PIV additional to those
described in
\S5 have been derived very recently and a discussion of these may be
found in \cite{\refBC}.

\section{\bts\ for the fourth \p\ equation}
The first \bt\ for PIV was derived by Lukashevich \cite{\refLukash} who
wrote
PIV (\eqPIV) in terms of a system of Riccati equations. Rather than
adopt
the method given by Lukashevich, we follow Gromak \cite{\refGromak} and
consider the system
$$\eqalignno{w' &=q+2 \eps z w+\eps w^2+2 \eps wv, \qquad
v' =p-2 \eps z v -\eps v^2-2 \eps wv, &\eqnm{eqqIIi}{a,b}\cr}$$
where $\eps^2=1$, $q^2=-2 \beta$ and $p=-1-\alpha \eps-\tfr12q$. If $v$
is eliminated between these equations then it is easily shown that $w$
satisfies PIV (\eqPIV). However, if $w$ is eliminated then it
transpires that $v$ also is a solution of PIV though not with the
parameters
$\alpha$ and $\beta$ as in (\eqPIV), but rather with $\alpha_1$ and
$\beta_1$
where
$\alpha_1=\tfr14 \left(2 \eps-2 \alpha +3 \eps \sqrt{-2\beta}\right)$
and
$\beta_1=-\tfr12 \left( 1+\alpha \eps +\tfr12 \sqrt{-2\beta}
\right)^2$.
As equation (\eqqIIi{a}) may be rewritten in the form
$$v= \frac{w'-2 \eps zw -\eps w^2-\sqrt{-2\beta}}{2\eps
w},\eqn{eqqIIiii}$$
it is apparent that if $w(z;\alpha,\beta)$ denotes a solution of PIV
for parameters $\alpha$ and $\beta$ then setting $\eps=\pm 1$ in
(\eqqIIi--\eqqIIiii) gives rise to two further solutions of PIV. If the
transformations from
$w(z;\alpha,\beta)$ to these new solutions are denoted by $\~W^\pm$ and
$\^W^\pm$ then these B\"acklund formulae can be written as
$$\eqalignno{
\~W^\pm(w(z;\alpha,\beta)):&=w(z;\~\alpha^\pm,\~\beta^\pm) =
{w'(z;\alpha,\beta) -
w^2(z;\alpha,\beta)-2zw(z;\alpha,\beta)\mp\sqrt{-2\beta}\over
2w(z;\alpha,\beta)},\qquad\qquad &\eqnm{eqqIIiv}{a}\cr
\~\alpha^\pm &= \tfr14\left(2-2\alpha \pm 3\sqrt{-2\beta}\,\right),
\quad\quad
\~\beta^\pm = -\tfr12\left(1+\alpha \pm
\tfr12\sqrt{-2\beta}\,\right)^{\! 2},
&\eqnr{b,c}\cr\cr
\^W^\pm(w(z;\alpha,\beta)):&=w(z;\^\alpha^\pm,\^\beta^\pm)=
-\,{w'(z;\alpha,\beta) +
w^2(z;\alpha,\beta)+2zw(z;\alpha,\beta)\mp\sqrt{-2\beta}\over
2w(z;\alpha,\beta)},\qquad\qquad &\eqnm{eqqIIv}{a}\cr
\^\alpha^\pm&= -\tfr14\left(2+2\alpha \pm
3\sqrt{-2\beta}\,\right),\quad\quad
\^\beta^\pm = -\tfr12\left(1-\alpha \pm
\tfr12\sqrt{-2\beta}\,\right)^{\! 2}.
&\eqnr{b,c}\cr}$$ Thus given a solution $w(z;\alpha,\beta)\not\equiv 0$
of PIV,
the solutions
$w(z;\~\alpha^\pm,\~\beta^\pm)$  and $w(z;\^\alpha^\pm,\^\beta^\pm)$
may be
obtained using (\eqqIIiv) and (\eqqIIv). Moreover, the transformations
(\eqqIIiv) and (\eqqIIv) are effectively nonlinear differentiation
formulae for
solutions of PIV since $w(z;\~\alpha^\pm,\~\beta^\pm)$  and
$w(z;\^\alpha^\pm,\^\beta^\pm)$ are expressed in terms of
$w(z;\alpha,\beta)$
and its derivatives.

Kitaev \cite{\refKitpc} derived two further sets of \bts\ for PIV using
the
associated iso\-mono\-dromy deformation representation, which has a
regular singular
point at $\lambda=0$ and an irregular singular point of rank 2 at
$\lambda=\infty$ (cf., \cite{\refFMA}). Kitaev showed that the
\bts\ $\~W$ and
$\^W$ correspond to the regular point at $\lambda=0$ and the
transformations
$W^{\sdag}$ and $W^{\sddag}$, defined below, are associated with the
irregular
point at $\lambda=\infty$. If $w(z;\alpha,\beta)$ is a solution of PIV,
then so
also are $w^{\sdag}(z;\alpha^{\sdag},\beta^{\sdag})$ and
$w^{\sddag}(z;\alpha^{\sddag},\beta^{\sddag})$ where
$$\eqalignno{W^{\sdag\pm}(w(z;\alpha,\beta)):&=
w^{\sdag\pm}(z;\alpha^{\sdag\pm},\beta^{\sdag\pm})
\cr&=w(z;\alpha,\beta)+{2\left(1-\alpha\mp\tfr12\sqrt{-2\beta}\,\right)
w(z;\alpha,\beta)\over
w'(z;\alpha,\beta)\pm\sqrt{-2\beta}+2zw(z;\alpha,\beta)+w^2(z;\alpha,
\beta)},&\eqnm{eqqIIxxi}{a}\cr
 \alpha^{\sdag\pm}&=\tfr32-\tfr12\alpha\mp\tfr34\sqrt{-2\beta},\quad\quad
\beta^{\sdag\pm}=-\tfr12\left(1-\alpha\pm\tfr12\sqrt{-2\beta}\,\right)^{\!
2},
&\eqnr{b,c}\cr
W^{\sddag\pm}(w(z;\alpha,\beta)):&=
w^{\sddag\pm}(z;\alpha^{\sddag\pm},\beta^{\sddag\pm}) \cr&=
w(z;\alpha,\beta)+{2\left(1+\alpha\pm\tfr12\sqrt{-2\beta}\,\right)
w(z;\alpha,\beta)\over
w'(z;\alpha,\beta)\mp\sqrt{-2\beta}-2zw(z;\alpha,\beta)-w^2(z;\alpha,
\beta)},&\eqnm{eqqIIxxii}{a}\cr
 \alpha^{\sddag\pm}&=-\tfr32-\tfr12\alpha\mp\tfr34\sqrt{-2\beta},\quad\quad
\beta^{\sddag\pm}=-\tfr12\left(-1-\alpha\pm\tfr12\sqrt{-2\beta}\,\right)^{\!
2}.
&\eqnr{b,c}\cr}$$
These two transformations are valid for all solutions
$w(z;\alpha,\beta)$ for
which the numerators and denominators are non-zero. The transformation
$W^{\sdag\pm}$ is equivalent to the transformations
$T_1$ (lower sign) and $T_2$ (upper sign) given by Murata
\cite{\refMurata}.
In \cite{\refBCHI}, is was shown that the \bts\ $W^{\sdag\pm}$ and
$W^{\sddag\pm}$ are expressible in terms of $\~W^\pm$ and $\^W^\pm$
$$\eqalignno{
w^{\sdag\pm}(z;\alpha^{\sdag\pm},\beta^{\sdag\pm}) &= \cases{
\~W^+\^W^\mp(w(z;\alpha,\beta)), & if
$1-\alpha-\tfr12\sqb>0$,\cr \~W^-\^W^\mp(w(z;\alpha,\beta)), &
if $1-\alpha-\tfr12\sqb<0$,\cr}\cr
w^{\sddag\pm}(z;\alpha^{\sddag\pm},\beta^{\sddag\pm}) &=\cases{
\^W^+\~W^\pm(w(z;\alpha,\beta)), & if
$1+\alpha+\tfr12\sqb>0$,\cr \^W^-\~W^\pm(w(z;\alpha,\beta)), &
 if $1+\alpha+\tfr12\sqb<0$.\cr}\cr}$$
Consequently the following {\it Kitaev fractional transformations} can
be
derived
$$\eqalignno{
w^{\sdag\pm}(z;\alpha^{\sdag\pm},\beta^{\sdag +})&=w(z;\alpha,\beta)
-{1-\alpha\mp\tfr12\sqb\over \^W^\mp(w(z;\alpha,\beta)) },\cr
w^{\sddag\pm}(z;\alpha^{\sddag\pm},\beta^{\sddag\pm})&=w(z;\alpha,\beta)
+{1+\alpha\pm\tfr12\sqb\over \~W^\pm(w(z;\alpha,\beta)) },
\cr}$$
which are three-term recurrence relations, or {\it nonlinear
superposition formulae}, for PIV (see \cite{\refBCHI} for further
details.
These results can be used for the efficient derivation of solution
hierarchies
of PIV by use of {\it algebraic\/} manipulations alone.

Fokas \etal \cite{\refFMA} who, using a Schlesinger transformation
formulation
related to the iso\-mono\-dromy method, also deduced a number of
\bts\ for PIV.
However, as for the \bts\ $W^{\sdag\pm}$ and  $W^{\sddag\pm}$ above,
these are
expressible in terms of $\~W^\pm$ and $\^W^\pm$ \cite{\refBCHI}.

\section{Solution Hierarchies for the fourth \p\ equation}
\subsection{Introduction}
In this section we shall discuss various one-parameter families
of solutions and rational solutions for PIV. In order to characterise
such
families we start by recalling that the Riccati system associated with
PIV is
given by (\eqqIIi). Eliminating either $v$ or $w$ yields PIV.
If $v\equiv 0$ in (\eqqIIi) then $p\equiv 0$ so that
$q =-2 (\alpha\eps+1)$ and hence
$$\beta =-2 (1+\alpha\eps)^2,\eqn{eqbetop}$$
which we shall refer to as the {\it one-parameter family condition}.
One-parameter solutions are found by solving the Riccati equation,
$$w' -2 \eps z w-\eps w^2-2(1+\alpha\eps)=0. \eqn{eqopre}$$
Applying the linearizing transformation $w=-\eps u'/u$ to this form
yields
the Weber-Hermite equation
$$u''-2\eps zu'-2(\alpha+\eps)u=0. \eqn{eqwhe}$$
Further, if we make the transformation
$u(z)=\eta(\xi)\exp\left(\tfr14\eps
\xi^2\right)$, with $\xi=\sqrt2\,z$, then we obtain the parabolic
cylinder
function equation (cf., \cite{\refASte})
$${\d^2\eta\over\d\xi^2}=\left(\tfr14\xi^2-\nu-\tfr12\right)\eta,\eqn{eqpcf}$$
with $\nu=-\alpha-\tfr12(1+\eps)$. If $\nu\not\in\Z^+$, then this
equation has
the general solution
$$\eta(\xi)=AD_{\nu}(\xi)+BD_{\nu}(-\xi),$$
where $A$ and $B$ are arbitrary
constants and $D_{\nu}(\xi)$ is the parabolic cylinder function (cf.,
\cite{\refASte}). If $\nu\in\Z^+$, then
$D_n(\xi)=\He_n(\xi)\exp\left(-\tfr14\xi^2\right)$, where $\He_n(\xi)$
is the
Hermite polynomial of degree $n$ given by
$$\He_n(\xi)=(-1)^n\exp\left(\tfr12 \xi^2\right){\d^n\over\d \xi^n}
\left[\exp\left(-\tfr12
\xi^2\right)\right].\eqn{hermite}$$

In general, the one-parameter families of solutions for PIV expressible
in terms of parabolic
cylinder functions are given by
$$\eqalignno{
&{\hbox to 150pt{$\displaystyle w(z;-(\nu+1),-2\nu^2)=-{\phi_{\nu}'(z)
\over\phi_{\nu}(z)}$,\hfill}}
\qquad
w(z;-\nu,-2(\nu+1)^2)=-2z-{\phi_{\nu}'(z)\over\phi_{\nu}(z)},&\eqnm{eqXVI}{a,b}\cr
&{\hbox to 150pt{$\displaystyle w(z;\nu+1,-2\nu^2)={\psi_{\nu}'(z)\over
\psi_{\nu}(z)}$,\hfill}}
\qquad w(z;\nu,-2(\nu+1)^2)=-2z-{\psi_{\nu}'(z)\over\psi_{\nu}(z)},
&\eqnr{c,d}\cr}$$ where $\phi_{\nu}(z)$ satisfies the Weber-Hermite
equation
(\eqwhe) with $\eps=1$ and
$\alpha=-\nu-1$ and $\psi_{\nu}(z)$ satisfies (\eqwhe) with $\eps=-1$
and
$\alpha=\nu+1$ (cf., \cite{\refPAC,\refLukash}). If $\nu=n$, with $n$ a
positive
integer, then
$\phi_n(z)$ and $\psi_n(z)$ are polynomials of degree $n$ expressible
in terms
of the Hermite polynomial $\He_n(\xi)$ (\hermite).

\subsection{The Complementary Error Function Hierarchy}
If the parameters in (\eqopre) are $\alpha=1$ and $\eps=-1$, then
(\eqwhe) may
be integrated to yield $$u(z)=B-A\erfc (z),\eqn{}$$
where $A$ and $B$ are arbitrary constants and $\erfc (z)$ is the
complementary error function given by
$$\erfc (z)=\fr{2}{\sqrt{\pi}} \int_z^\infty \exp (-t^2) \,\d
t.\eqn{deferf}$$
Hence we obtain the exact solution of PIV
$$w(z;1,0)=\Psi(z)\equiv\fr{2A \exp
(-z^2)}{\sqrt{\pi}\,\left[B-A\erfc(z)\right]}
\eqn{eqIIv}$$ (see also \cite{\refGromak}).

Applying the \bts\ $\~W^{\pm}$, $\^W^{\pm}$, $W^{\sddag\pm}$, and
$W^{\sddag\pm}$ to (\eqIIv) yields a hierarchy of solutions of PIV of
the form
$$w(z;m_1+1,-2m_2^2)={R(z,\Psi(z))},$$
where $m_1$ and $m_2$ are integers such that $\tfr12(m_1+m_2)\in\Z^+$
(i.e.,
either $m_1$ and $m_2$ are either both even integers or both odd
integers), and
$R(z,\Psi(z))$ is a rational function of its arguments; see Tables
3.2.2 and 3.2.3 in \cite{\refBCHI} for further details. For example,
applying
the transformation $W^{\sddag +}$ to (\eqIIv) yields
$$W^{\sddag +}\left(w(z;1,0)\right)=w(z;-2,-2)=w(z;1,0)+{2\over
\~W^+\left(w(z;1,0)\right)},$$ and so since
$$\~W^+\left(w(z;1,0)\right)=w(z;0,-2)=-2z-w(z;1,0)\equiv-2z-\Psi(z)$$
then
$$w(z;-2,-2)=\Psi(z)-{2\over 2z+\Psi(z)}.$$

We remark that the subset of this hierarchy given by $w(z;2n+1,0)$
correspond
to the bound state solutions discovered in \cite{\refBCHM} (see also
\cite{\refBCH}), which have the property that they decay exponentially
as
$z\to\pm\infty$ and are analogues of the bound state solutions for the
linear
harmonic oscillator (cf., \cite{\refSchiff}).

\subsection{The Half-integer Hierarchy}
If the parameters in (\eqopre) are $\alpha=\pm\tfr12$ and
$\beta=-\tfr12$, then
under the transformation $w(z)=\pm{\theta'_\pm(z)/\theta_\pm(z)}$
with $\theta_\pm(z)=\eta(\xi)\,\exp\left(\mp\tfr14\xi^2\right)$ and
$\xi=\sqrt{2}\,z$, it follows that $\eta(\xi)$ satisfies (\eqpcf) with
$\nu=-\tfr12$. Hence $\eta(\xi)= AD_{-1/2}(\xi)+ BD_{-1/2}(-\xi)$, with
$A$ and
$B$ arbitrary constants,  and so we obtain the following solutions of
PIV
$$w(z;\pm\tfr12,-\tfr12)=\fr{\theta_\pm'(z)}{\theta_\pm(z)}= -z \pm
\Theta(z),\eqn{eqhint}$$ where
$$\Theta(z) = -z+\fr{\sqrt{2}\,\left[AD_{1/2}(\sqrt{2}\,z)-
BD_{1/2}(-\sqrt{2}\,z)\right]}{AD_{-1/2}(\sqrt{2}\,z)+BD_{-1/2}(-\sqrt{2}\,z)}.
\eqn{eqTheta}$$ Applying the \bts\ $\~W^\pm$ and $\^W^\pm$ (\eqqIIiv,
\eqqIIv)
to $w(z;\pm\tfr12,-\tfr12)$ yields a hierarchy of solutions of PIV
which have
importance in connection with quantum gravity (see \S4 below for
further
details).

\subsection{Rational Solution Hierarchies}
It is easily verified that $w=1/z$ satisfies PIV (\eqPIV) with the
parameter
choices $\alpha=2$, $\beta=-2$ and this is a simple example of a
rational
solution of PIV. Two families of rational solutions for PIV take the
forms
$$w(z;\alpha,\beta)={P_{n-1}(z)\over Q_n(z)},\qquad\qquad
 w(z;\alpha,\beta)=-2z+{P_{n-1}(z)\over Q_n(z)},\eqno\eqnm{eqIV}{a,b}$$
where $P_m(z)$ and $Q_m(z)$ denote some polynomials of degree $m$
consisting
of either entirely even or else entirely odd powers of $z$. Murata
\cite{\refMurata} has shown that (\eqPIV) admits unique rational
solutions of
type (\eqIV{a}) or (\eqIV{b}) if the parameters are of the form
$$\eqalignno{(\alpha,\beta)&=\left( \pm k,-2(1+2n+k)^2\right);\qquad k,
n\in\Z,
\quad n\leq -1,\quad k\geq -2n,&\eqnm{eqV}{a}\cr
(\alpha,\beta)&=\left( k, -2(1+2n+k)^2\right);\qquad k, n\in\Z,\quad
n\geq 0,\quad k\geq -n,&\eqnr{b}\cr}$$
respectively. A third family of rational solutions of PIV is
characterised by
$$w(z;\alpha,\beta)=-\tfr23z+{P_{n-1}(z)\over Q_n(z)},\eqn{eqVI}$$
where $(\alpha,\beta)=\left( n_1,-\tfr29(1+3n_2)^2\right)$, with $n_1$
and $n_2$
either both even or both odd integers. An extensive  discussion of the
important
properties of these rational solution families together with tables
containing
the first few solutions in each of these three hierarchies is given in
\cite{\refBCHI}.

\def\wnt{w_{n,t}}%{\d w_n\over\d t}
\section{Exact solutions of the discrete first \p\ equation}
It is well known that the Korteweg-de Vries (KdV) equation, which
possesses
a similarity reduction to PI, has several discrete versions; one of
these
is the Kac-Moerbeke equation
$$\wnt=-w_n(w_{n+1}- w_{n-1}),\eqn{eqVIIi}$$
which reduces to the KdV equation in the appropriate continuous limit.
Fokas
\etal \cite{\refFIK} have demonstrated that in the case of (\eqVIIi) a
similarity  solution is characterised by solving it simultaneously with
the
discrete equation
$$n=2tw_n+w_n(w_{n+1}+w_n+w_{n-1})\eqn{eqVIIii}$$
and the continuous limit $w_n=-\tfr23t(1-2\eps^2\eta(z))$,
$n=-\tfr29t^2\left(1+z\eps^4\right)$, as $\eps\to 0$ maps (\eqVIIii) to
PI
(see \cite{\refGM}). Thus equation (\eqVIIii) is a version of discrete
PI and
it has arisen in the theory of two-dimensional quantum gravity (cf.,
\cite{\refFIK,\refNP}). We can use (\eqVIIi) to substitute for
$w_nw_{n+1}$ and
$w_nw_{n-1}$ in (\eqVIIii) and these two equations can be rewritten as
$$w_{n+1}={n-\wnt-2tw_n-w_n^2\over 2w_n},\qquad
  w_{n-1}={n+\wnt-2tw_n-w_n^2\over 2w_n}.\eqno\eqnm{eqVIIiv}{a,b}$$
We observe that (\eqVIIiv) is essentially the same Riccati system as
(\eqqIIi)
since (\eqVIIiv{a}) becomes (\eqqIIi{a}) when $\eps=1$ and $w_n$,
$w_{n+1}$, $n$
and $t$ are replaced by $v$, $w$, $p$  and $z$ respectively and
(\eqVIIiv{b})
becomes (\eqqIIi{b}) if $w_n$, $w_{n-1}$, $n$ and
$t$ are identified with $w$, $v$, $-q$ and $z$, respectively. If $n$ is
replaced by $n+1$ in
(\eqVIIiv{b}) and $w_{n+1}$ eliminated by using (\eqVIIiv{a}) it
follows that
$w_n(t)$ satisfies
$$w_nw_{n,tt}={\tfr12}\left(\wnt\right)^2+\tfr32w_n^4+4tw_n^3+2\left(t^2+
\tfr12n\right)w_n^2-\tfr12{n^2},\eqn{eqVIIvi}$$
which is precisely PIV with $\alpha=-\tfr12n$ and $\beta=-\tfr12 n^2$.
Thus the
solution of d-PI (\eqVIIvi) can be expressed in terms of solutions of
PIV with
parameters $\alpha=-\tfr12 n$,
$\beta=-\tfr12 n^2$ and $n\in\Z$. In \cite{\refBCHI} we showed that the
first
few solutions  of (\eqVIIvi), and hence also (\eqVIIi) and (\eqVIIii),
are
$$\eqalignno{
w_{\pm1}(t)&=w(t;\pm\tfr12,-\tfr12)=
-t\mp\Theta(t),&\eqnm{eqVIIix}{a}\cr
w_{\pm2}(t)&=w(t;\pm1,-2)= \pm{\Theta^2(t)-t^2\pm 1\over \Theta(t)\pm
t},
&\eqnr{b}\cr w_{\pm3}(t)&=w(t;\pm\tfr32,-\tfr92)= -{ 3\Theta^2(t)\pm
4t\Theta(t)+t^2\pm 1\over [\Theta(t)
\pm t][\Theta^2(t)-t^2\pm 1]}, &\eqnr{c}
\cr}$$
with $\Theta(t)$ as given in (\eqTheta). We remark that Fokas \etal
\cite{\refFIK} obtained the solution $w_1(t)$ by solving an associated
Riemann-Hilbert problem for PIV; the derivation given here is much
simpler.

\section{Rational solutions of the discrete fourth \p\ equation}
Tamizhmani \etal \cite{\refTGR} noted that if we set
$\eta=3\delta^{-2}+\gamma^2-2(a^2+b^2)$ and $\mu=a^2-b^2$,
then the d-PIV equation (\eqI) can be factorised as
$$(x_{n+1}+x_n)(x_{n-1}+x_n)={(x_n+a+b)(x_n+a-b)(x_n-a+b)(x_n-a-b)
\over (x_n+z_n+\gamma)(x_n+z_n-\gamma)},\eqn{eqVII}$$
and the parameters $\alpha=\tfr14\eta$ and $\beta=-\tfr12\mu^2$
are now given by
$$\alpha=\tfr34\delta^{-2}+\tfr14\gamma^2-\tfr12(a^2+b^2),\qquad
\beta=-\tfr12(a^2-b^2)^2.\eqn{eqVIII}$$
Simple rational solutions of (\eqVII) can be found with $x_n$
proportional to $z_n$; these are
$$\eqalignno{ x_n&=-2z_n,\qquad a=\tfr12\delta+\gamma,\quad
 b=\tfr12\delta-\gamma,
&\eqnn{eqIXa}\cr
x_n&=-\tfr23z_n,\qquad a=\tfr16\delta+\gamma,\quad
b=\tfr16\delta-\gamma.
&\eqnn{eqIXb}\cr}$$
In the limit as $\delta\to 0$, with $\gamma=1/\delta$, these discrete
solutions tend to  $w(z;0,-2)=-2z$ and $w(z;0,-\tfr29)=-\tfr23z$,
respectively, which are the first members
of the PIV hierarchies typified by (\eqIV{b}) and
(\eqVI). If solutions of (\eqVII) proportional to $1/z_n$ are sought
then
it is found that
$$x_n=-{\delta(\delta\pm\gamma)\over z_n},\qquad
a=\tfr32\delta\pm\gamma,
\quad b=\tfr12\delta\pm\gamma,\eqn{eqX}$$
and so, in the limit as  $\delta\to 0$, with $\gamma=1/\delta$,
we have $x_n\to w(z;\pm 2,-2)=\pm 1/z$ and these
continuous solutions are members of the family described by
(\eqIV{a}).

More complicated rational solutions of d-PIV (\eqI) can be deduced by
rewriting (\eqVII) as the pair
$$x_n+x_{n-1}={ (x_n+a+b)(x_n+a-b)\over (x_n+z_n\pm \gamma)},
\qquad x_n+x_{n+1}={ (x_n-a+b)(x_n-a-b)\over (x_n+z_n\mp \gamma)},
\eqno\eqnm{eqXI}{a,b}$$
which are  discrete analogues of the Riccati equation (\eqopre).
Tamizhmani \etal\ \cite{\refTGR} speculated that such a formalism ought
to lead
to discrete rational solutions though they did not present any such
solutions.
It is a routine calculation to verify that equations (\eqXI{a}) and
(\eqXI{b}) are compatible if and only if $a=\tfr12\delta\pm\gamma$.
[We remark at this stage that (\eqXI) is not the only possibility for
the
splitting of (\eqVII). Easy generalisations of (\eqXI) include the
multiplication of the right hand sides of (\eqXI{a}) and (\eqXI{b}) by
constants $C$ and $1/C$ respectively or the taking of the factors in
different
pairings. However, it can be shown that in either of these cases we
obtain an incompatible couple of equations so that the choice of
factorisation (\eqXI) is not as specialised and restrictive as it might
first
appear.]

If $a=\tfr12\delta\pm\gamma$, then we can seek solutions of (\eqI)
by finding solutions of the simpler form (\eqXI{b}) (or (\eqXI{a})
would
do equally well). If we write $x_n=P_n/Q_n$ then (\eqXI{b}) can be
recast
as
$$\eqalignno{&\kappa_{n+1}P_{n+1} =(z_n\pm\gamma+\delta)P_n-\mu Q_n,
&\eqnm{eqXIII}{a}\cr
       &\kappa_{n+1}Q_{n+1} =-(z_n\mp\gamma)Q_n-P_n, &\eqnr{b}\cr}$$
where the `separation' parameter $\kappa_{n+1}$ can depend both on $n$
and $z_n$ (recall that $\mu=a^2-b^2$). However, if for the moment
we take $\kappa_{n+1}=1$, then eliminating $P_n$ between (\eqXIII{a})
and
(\eqXIII{b}), setting $\gamma=1/\delta$ and taking the limit as
$\delta\to 0$
shows that $Q_n\to q(z)$ where $q(z)$ satisfies
$$q''=(z^2+\mu-1)q.\eqn{eqXIV}$$
If in this equation we set $q(z)=\eta(\xi)$ with $\xi=\sqrt2\,z$ then
we obtain
the parabolic cylinder equation (\eqpcf) with $\nu=-\tfr12\mu$.

If the variable $Q_n$ is eliminated from equations (\eqXIII) and the
usual limit taken then it follows in a manner similar to that outlined
above
that rational solutions $x_n=P_n/Q_n$ of (\eqI) exist which tend to a
function
of the form $-\He_m'(\xi)/\He_m(\xi)$ whenever $\mu=-2m$, with
$m\in\Na$.
As illustrated in \S3.1 above, solutions of PIV (\eqPIV) can be
expressed in
terms of parabolic cylinder functions when the parameters $\alpha$ and
$\beta$
take certain values.  Motivated by these comments concerning the
(continuous)
PIV case, we return to equations (\eqXIII) for the discrete situation.
If we
write
$$(P_n,Q_n)=(A_n,B_n)\times (-\delta)^n\Gamma\left( {z_n-m\delta
\mp\gamma\over\delta}\right),\eqn{eqXVIII}$$
where $\Gamma(z)$ denotes the usual Gamma function, and choose
$\kappa_{n+1}\equiv 1$, then we obtain the pair
$$\eqalignno{
   &(z_n-m\delta\mp\gamma)A_{n+1}+(z_n+\delta\pm\gamma)A_n=\mu B_n,
&\eqnm{eqXIX}{a}\cr
   &(z_n-m\delta\mp\gamma)B_{n+1}-(z_n\mp\gamma)B_n=A_n.&\eqnr{b}\cr}$$
For $m\in\N$ we can find exact solutions of these equations with
$A_n$ and $B_n$ taking the forms of polynomials in $z_n$, consisting
of either only even or only odd powers, and of degrees $m-1$ and $m$
respectively. The first few solutions in this hierarchy are
\def\de{\delta}
\def\ga{\gamma}
$$\eqalignno{
   m=1,\qquad&\qquad x_n=-{\delta(\delta\pm\gamma)\over
   z_n},&\eqnm{eqXX}{a}\cr
   m=2,\qquad&\qquad
   x_n=-{2\de(3\de\pm2\ga)z_n\over2z_n^2-\de(2\de\pm\ga)},
&\eqnr{b}\cr
   m=3,\qquad&\qquad x_n=-{3\de(2\de\pm\ga)\left[2z_n^2-\de
(3\de\pm\ga)\right]\over z_n\left[ 2z_n^2-\de(8\de\pm 3\ga)\right] },
&\eqnr{c}\cr
   m=4,\qquad&\qquad x_n=-{4\de(5\de\pm 2\ga)\left[ 2z_n^2-\de
(11\de\pm 3\ga)\right]z_n\over 4z_n^4-4\de(10\de\pm 3\ga)z_n^2+3\de^2
(3\de\pm\ga)(4\de\pm\ga)}.
&\eqnr{d}\cr
   m=5,\qquad&\qquad x_n=-{5\de(3\de\pm\ga)\left[ 4z_n^4-4\de(13\de
\pm 3\ga)z_n^2+3\de^2(3\de\pm\ga)(4\de\pm\ga)\right]\over
z_n\left[ 4z_n^4-20\de(4\de\pm\ga)z_n^2+\de^2(256\de^2\pm
125\ga\de+15\ga^2)
\right] }.&\eqnr{e}
\cr}$$

In each case the corresponding value of $\mu$ in (\eqXIX{a}) is
$\mu=-\de m\left[ (m+1)\de \pm 2\ga\right]$ and this leads to the
respective
values of parameters $\alpha$ and $\beta$, as defined by (\eqVIII),
given by
$$\alpha=\tfr34(\de^{-2}-\ga^2)\mp
(m+1)\ga\de-\tfr14\de^2\left[2m(m+1)+1
\right],
\qquad \beta=-\tfr12m^2\delta^2\left[ (m+1)\de\pm 2\ga\right]^2.
\eqn{eqCVS}$$
We emphasise at this stage that the discrete solutions (\eqXX) are
exact and are valid for {\it any} $\de$ and $\ga$. We can recover
continuous
solutions by letting $\ga=1/\de$ and taking the limit as $\de\to 0$
which yields
$\alpha=\mp(m+1)$ and $\beta=-2m^2$. Then from (\eqXX) we obtain
$$\eqalign{
&w(z;\mp 2,-2)=\mp {1\over z},\qquad
w(z;\mp 3,-8)=\mp {4z\over 2z^2\mp 1},\qquad
w(z;\mp 4,-18)= \mp {3(2z^2\mp 1)\over z(2z^2\mp 3)},\cr
&w(z;\mp 5,-32)= \mp {8z(2z^2\mp 3)\over 4z^4\mp 12z^2+3},\qquad
w(z;\mp 6,-50)=\mp {5(4z^4\mp 12z^2+3)\over z(4z^4\mp
20z^2+15)}.\cr}$$
It is then clear that solutions (\eqXX) can be
thought of as discrete analogues of the (continuous) solutions of PIV
taking
forms given by (\eqXVI{a}) and (\eqXVI{d}), with $\nu=m$.

If in (\eqXIII) we let
$$(P_n,Q_n)=(A_n,B_n)\times \delta^n\thinspace\Gamma\left( {z_n-m\delta
\pm\gamma\over\delta}\right),\eqn{eqXXII}$$
then we obtain
$$\eqalignno{
   &(z_n-m\delta\pm\gamma)(A_{n+1}-A_n)=-\mu B_n,
&\eqnm{eqXXIII}{a}\cr
   &(z_n-m\delta\pm\gamma)B_{n+1}+(z_n\mp\gamma)B_n=-A_n.&\eqnr{b}\cr}$$
As previously, exact polynomial solutions of these can be derived for
integer
$m$ except that now $A_n$ and $B_n$ are of degrees $m+1$ and $m$
respectively.
Then the first few solutions of this type are
$$\eqalignno{ m=0,\qquad&\qquad x_n=-2z_n,&\eqnm{eqXXIV}{a}\cr
       m=1,\qquad&\qquad x_n=-{2z_n^2-\de(\de\mp\ga)\over z_n},
&\eqnr{b}\cr
       m=2,\qquad&\qquad x_n=-{2z_n\left[2z_n^2-\de(5\de\mp
3\ga)\right]\over 2z_n^2-\de(2\de\mp\ga)},
&\eqnr{c}\cr
       m=3,\qquad&\qquad x_n=-{4z_n^4-4\de(7\de\mp 3\ga)z_n^2
+3\de^2(3\de\mp\ga)(2\de\mp\ga)\over z_n\left[ 2z_n^2-\de(8\de\mp
3\ga)\right] },&\eqnr{d}\cr
       m=4,\qquad&\qquad x_n=-{2z_n\left[ 4z_n^4-20\de
(3\de\mp\ga)z_n^2+\de^2(146\de^2\mp 95\ga\de+15\ga^2)\right] \over
4z_n^4-4\de(10\de\mp 3\ga)z_n^2+3\de^2(4\de\mp\ga)(3\de\mp\ga)}.
&\eqnr{e}\cr}$$
The corresponding values of the parameters $\alpha$ and $\beta$ are
given
by
$$\alpha=\tfr34(\de^{-2}-\ga^2)\pm m\ga\de-\tfr14\de^2\left[2m(m+1)+
1\right],\qquad
 \beta=-\tfr12(m+1)^2\de^2(m\de\mp 2\ga)^2\eqn{eqasasd}$$
and in the limit $\de\to 0$ with $\ga=1/\de$,
solutions (\eqXXIV) reduce to the
functions
$$\eqalignno{
&w(z;0,-2) =-2z,\qquad
w(z;\pm 1,-8)=-2z\mp{1\over z},\qquad
w(z;\pm 2,-18)=-2z\mp{4z\over 2z^2\pm 1},\cr
&w(z;\pm 3,-32)=-2z\mp{3(2z^2\pm 1)\over z(2z^2\pm 3)},\qquad
w(z;\pm 4,-50)=-2z\mp{8z(2z^2\pm 3)\over 4z^4\pm 12z^2
+3}.\cr}$$
These are all members of the so-called `$-2z$' hierarchy of rational
solutions for PIV (\eqPIV) and are of the form (\eqXVI{b}) or
(\eqXVI{c}), with $\nu=m$. It is straightforward to obtain further
exact rational solutions of d-PIV by
solving the pairs (\eqXIX) or (\eqXXIII) for higher values of $m$.

We note here that although we have found some rational solutions of
d-PIV (\eqI) it can be anticipated that further such solutions exist
which
are not derivable using the procedure described above. This deduction
follows from the observation that rational solutions of PIV (\eqPIV)
 are possible for all parameter values as described by
(\eqV) and that so far we have discrete solutions which, in the
appropriate limit, tend to the continuous solutions of type (\eqXVI).
Therefore, there should be discrete counterparts to the remaining
continuous rational solutions. These can be constructed by appealing to
some \bts\ for d-PIV which were given by Tamizhmani \etal\
\cite{\refTGR}. These authors presented a sequence of transformations
which, given a solution of (\eqVII) with parameters $a$, $b$ and
$\gamma$,
yields a further solution of the same equation but now with parameters
$a+\delta$, $b$ and $\gamma$.
In short, suppose that $x_n$ is a solution of d-PIV  with
parameters $a$, $b$ and $\gamma$. Then
$$\~x_n=-{x_nx_{n+1}+x_n(\~z_n+a)+x_n(\~z_n-a)+b^2-a^2\over
x_n+x_{n+1}},\eqn{eqback}$$
with $\~z_n= z_n+\tfr12\delta$ also satisfies d-PIV but now with
parameters $\~a=a+\tfr12\delta$, $\~b=\gamma$ and $\~\gamma=b$.
The subtlety with transformation (\eqback) is that $\~x_n$ is defined
on
a lattice of points which is offset by $\tfr12\delta$ from the
original. In
order to obtain a solution valid at points coinciding with the initial
lattice
it is therefore necessary to reapply (\eqback) which has the overall
effect of
raising the value of $a$  by $\delta$ whilst keeping
$b$ and $\gamma$ invariant. Clearly $M$ applications of this sequence
will increment $a$ to $a+M\delta$ and leaves the other two parameters
unchanged so that from a starting solution with parameters
$a=\tfr12\de\pm\ga$, $b=(N+\tfr12)\de\pm\ga,$ for $N=0,1,2,\dots$
(which are the parameter values in (\eqI) appropriate to the family of
solutions whose first few members are listed in (\eqXX)) we can obtain
a solution with $a=(M+\tfr12)\de\pm\ga$ and $b=(N+\tfr12)\de\pm\ga$,
or, in terms of the parameters $\alpha$ and $\beta$ as given by
(\eqVIII),
$$\eqalignno{
 \alpha&=\tfr34(\de^{-2}-\ga^2)\mp\ga\de(M+N+1)+\left[ \tfr12(M-N)
(M+N+1)-(M+\tfr12)^2\right]\delta^2,&\eqnm{eqXXVIII}{a}\cr
 \beta&=-\tfr12(M-N)^2\de^2\left[(M+N+1)\de\pm 2\ga\right]^2.
&\eqnr{b}\cr}$$

In the usual limit our discrete rational solutions will then tend to
continuous ones with associated parameters $\alpha=\mp(M+N+1)$,
$\beta=-2(M-N)^2$. Now $M$ and $N$ can be chosen so as to force these
parameters to coincide with the form (\eqV{a}) for any $k$ and $n$ in
the permissible ranges; in this way we have a mechanism for deducing
discrete analogues of all those rational solutions of PIV (\eqPIV)
which take the form $P_{n-1}(z)/Q_n(z)$.

A similar argument can be applied to the discrete solutions (\eqXXIV)
and this demonstrates that given these results then suitable
application
of the \bts\ contained in \cite{\refTGR} will generate another
set of exact solutions of (\eqI). These comprise the discrete
counterpart to the `$-2z$' rational hierarchy of
(\eqPIV) which itself is characterised by the parameter values given in
(\eqV{b}).

We remark here that of the three hierarchies of rational solutions for
PIV (\eqPIV), the simple procedures outlined above have yielded
discrete
analogues of only two of these; no solutions corresponding to the
$-\tfr23z$
family (\eqVI) have been found. The reason for this is that although
the full
discrete equation (\eqVII) admits the solution $x_n=-\tfr23z_n$ with
$a=\tfr16\delta\pm\gamma$, the splitting of this equation into the two
Riccati-like forms (\eqXI) gives a pair of equations whose
compatibility
requires that $a=\tfr12\de\pm\ga$. Thus, as noted by Tamizhmani \etal
\cite{\refTGR}, this solution is not linearizable through the splitting
assumption and thus it is unsurprising that the discrete analogue of
the
$-\tfr23z$ hierarchy of solutions cannot be generated in this way.
However, we
shall now show how use of the \bt\ (\eqback) discussed above can be
used to
increment the values of $a$ or $b$ by
$\pm\delta$ and thus lead to some more exact solutions in the discrete
$-\tfr23z$ hierarchy. (It is noted that d-PIV (\eqVII) is invariant by
interchange of parameters $a$ and $b$ or by a sign change of either of
these.
Therefore, given that two applications of (\eqback) increases $a$ by
$\delta$
it is easy to see how a suitable changes in sign of $a$ and $b$ or the
interchanging
of these parameters allows transformations to be found that increase or
decrease the values of $a$ or $b$ by integral multiples of
$\delta$.) A few examples of simpler solutions in this hierarchy are:
$$\eqalignno{
x_n&=-\tfr23z_n,\qquad
a=\tfr16\delta\pm\gamma,\quad
b=-\tfr16\delta\pm\gamma,&\eqnm{eqtry}{a}\cr
x_n&=-\tfr23z_n-{\delta(\delta\mp3\gamma)\over 3z_n},\qquad
a=\pm\gamma-\tfr56\delta,\quad b=\mp\gamma+\tfr16\delta,&\eqnr{b}\cr
x_n&=-{2z_n(2z_n^2\mp 3\gamma\delta+\delta^2)\over 3(2z_n^2\pm 3\gamma
\delta-2\delta^2)},\qquad
a=\pm\gamma-\tfr76\delta,\quad b=\mp\gamma+\tfr56\delta,&\eqnr{c}\cr
  x_n&=-{2z_n(4z_n^4- 45\gamma^2\delta^2+\delta^4)\over
  3(2z_n^2-3\gamma
\delta-\delta^2)(2z_n^2+3\gamma\delta-\delta^2)},\qquad
a=\pm\gamma+\tfr56\delta,\quad  b=\mp\gamma+\tfr56\delta,&\eqnr{d}\cr
x_n&=-{ 8z_n^6-4\delta(\delta\pm3\gamma)z_n^4-2\delta^2(38\delta^2\mp
69
\gamma\delta+27\gamma^2)z_n^2+9\delta^3(4\delta^3\mp
11\gamma\delta^2+10
\gamma^2\delta\mp 3\gamma^3)\over 3z_n\left[4z_n^4-4\delta(4\delta\mp
3\gamma)z_n^2\pm 9\gamma\delta^2(\delta\mp\gamma)\right] },\cr
a&=\pm\gamma-\tfr76\delta,\quad
b=\mp\gamma+\tfr{11}6\delta.&\eqnr{e}\cr}$$
In the limit as $\delta\to 0$, with $\gamma\delta= 1$, the solutions
(\eqtry)
reduce to
$$\eqalignno{
&w(z;0,-\tfr29)=-\tfr23z,\qquad
 w(z;\pm1,-\tfr89)=-\tfr23z\mp{1\over z},\qquad
 w(z;\pm2,-\tfr29)=-\tfr23z\pm{4z\over 2z^2\pm 3},\cr
&w(z;0,-\tfr{50}9)=-\tfr23z\pm{24z\over (2z^2-3)(2z^2+3)},\qquad
 w(z;\pm3,-\tfr89)=-\tfr23z\pm{3(4z^4\pm 4z^2+3)\over z(4z^4\pm
 12z^2-9)},\cr}$$
which belong to the $-\tfr23z$ hierarchy of rational solutions of PIV
(\eqPIV);
see Table 4.1.3 of \cite{\refBCHI} for a more extensive list of such
solutions.
The method of generating the $-\tfr23z_n$ discrete rational solutions
given
in (\eqtry) is intrinsically less satisfying than that employed for the
evaluation of solutions (\eqXX) and (\eqXXIV) in the other two
families;
this feature is a direct consequence of the fact that the
$-\tfr23z_n$ forms do not arise from a suitable factorisation of d-PIV
into a
pair of simple compatible equations akin to (\eqXI). Instead the
present method
is based on a direct implementation of the \bt\ (\eqback), which, for
the for
more  complicated solutions in the hierarchy, becomes an increasingly
laborious
task. However, we believe that the \bt\ (\eqback) will inevitably be
needed to
make a complete evaluation of each of the hierarchies and provide a
systematic
and efficient procedure for doing this.

\ack
{%\baselineskip=10pt\rm
\rm The authors wish to thank Mark Ablowitz, Andrew Hicks, Liz
Mansfield, Bryce
McLeod and Alice  Milne for their helpful comments and illuminating
discussions.
PAC thanks the organisers for the invitation to the meeting and the
Royal
Society for a travel grant to attend it. The research of PAC is
partially
supported through U.K.\ Engineering and Physical Science Research
Council grant
GR/H39420 which is grateful acknowledged. This work was completed
whilst APB was
on leave at the School of Mathematics, University of New South Wales,
Sydney. He
is indebted to the Royal Society and the Australian Research Council
without
whose grants (the latter to Dr Peter Blennerhassett) his visit would
not have
been possible. In addition he is grateful to the staff and students of
New
College, UNSW for their provision of a Visiting Fellowship and to the
School of
Mathematics for their hospitality.}

\references{\frenchspacing
\baselineskip=11pt\parindent=20pt
\def\rm{\nrm}
\def\it{\nit}
\def\sl{\nsl}
\def\bf{\nbf}
\def\tt{\ntt}

\def\refbk#1#2#3#4#5{\item{\hbox to 15pt{\rm\expandafter \csname
#1\endcsname.\hfill}}\rm #2,\ {#3}, #4,\ #5.\par}
\def\refjl#1#2#3#4#5#6{\item{\hbox to 15pt{\rm\expandafter \csname
#1\endcsname.\hfill}}\rm #2,\ {\frenchspacing\it#3},\ {\bf#4}\ (#6),
#5.\par}
\def\refpp#1#2#3#4{\item{\hbox to 15pt{\rm\expandafter \csname
#1\endcsname.\hfill}}\rm #2,\ {#3}\ (#4).\par}
\def\refdpp#1#2#3#4{\item{\hbox to 15pt{\rm\expandafter \csname
#1\endcsname.\hfill}}\rm #2,\ {#3}\ (#4).\par}
\def\refdppp#1#2#3#4#5{\item{\hbox to 15pt{\rm\expandafter \csname
#1\endcsname.\hfill}}\rm #2,\ #4\ #3, {\it Exeter Preprint no.
#5}.\par}
\def\refcf#1#2#3#4#5#6#7{\item{\hbox to 15pt{\rm\expandafter \csname
#1\endcsname.\hfill}}\rm #2,\ #4 (#5), #6, #3, pp.\ #7.}
\rm

\refbk{refAC}{M.J. Ablowitz and P.A. Clarkson}{Solitons, Nonlinear
Evolution Equations and Inverse Scattering, {\frenchspacing\it
L.M.S.\ Lect.\
Notes Math.\/}, {\bf149}}{C.U.P., Cambridge}{1991}
\refjl{refASa}{M.J. Ablowitz and H. Segur}{Phys. Rev.
Lett.}{38}{1103--1106}{1977}
\refbk{refASte}{M. Abramowitz and I.A. Stegun}{Handbook of Mathematical
Functions}{Dover, New York}{1965}
\refjl{refBC}{A.P. Bassom and P.A. Clarkson}{Phys. Lett.
A}{194}{358--370}{1994}
\refjl{refBCH}{A.P. Bassom, P.A. Clarkson and A.C. Hicks}{IMA J. Appl.
Math.}{50}{167--193}{1993}
\refjl{refBCHI}{A.P. Bassom,\ P.A. Clarkson\ and A.C. Hicks}{Stud.
Appl. Math.}{}{to appear}{1995}
\refdpp{refBCHII}{A.P. Bassom,\ P.A. Clarkson\ and A.C. Hicks}
{``On the application of solutions of the fourth Painlev\'e equation to
various
physically motivated nonlinear partial differential
equations'', preprint {\bf M94/32}, Department of Mathematics,
University of
Exeter}{1994}
\refjl{refBCHM}{A.P. Bassom, P.A. Clarkson, A.C. Hicks and J.B.
McLeod}{Proc.
R. Soc. Lond. A}{437}{1--24}{1992}
\refjl{refBK}{E. Brezin and V. Kazakov}{Phys. Lett.
B}{236}{144--150}{1990}
\refjl{refPAC}{P.A. Clarkson}{Europ. J. Appl. Math.}{1}{279--300}{1990}
\refjl{refDoug}{M.R. Douglas}{Phys. Lett. B}{238}{176--180}{1990}
\refjl{refFGR}{A.S. Fokas, B. Grammaticos and A. Ramani}{J. Math.
Anal.
Appl.}{180}{342--360}{1993}
\refjl{refFIK}{A.S. Fokas, A.R. Its and A.V. Kitaev}{Commun. Math.
Phys.}{142}{313--344}{1991}
\refjl{refFMA}{A.S. Fokas, U.\ Mugan and M.J.
Ablowitz}{Physica}{30D}{247--283}{1988}
\refjl{refGNPRS}{B. Grammaticos, F.W. Nijhoff, V. Papageorgiou, A.
Ramani and J. Satsuma}{Phys. Lett. A}{185}{446--452}{1994}
\refcf{refGR}{B. Grammaticos and A. Ramani}{Kluwer, Dordrecht,
1993}{Applications of Analytic and Geometric Methods to Nonlinear
Differential
Equations}{P.A.\ Clarkson, ed.}{{NATO ASI Series C\/},
{\bf413}}{299--313}
\refjl{refRGV}{B. Grammaticos,\ A. Ramani and V. Papageorgiou}
{Phys. Rev. Lett.}{67}{1825--1828}{1991}
\refjl{refGromak}{V.I. Gromak}{Diff. Eqns.}{14}{1510--1513}{1977}
\refjl{refGM}{D.J. Gross and A.A. Migdal}{Phys. Rev.
Lett.}{64}{127--130}{1990}
\refbk{refInce}{E.L. Ince}{Ordinary Differential Equations}{Dover, New
York}{1956}
\refjl{refKOSGR}{K. Kajiwara, Y. Ohta, J. Satsuma, B. Grammaticos and
A.
Ramani}{J. Phys. A: Math. Gen.}{27}{915--922}{1994}
\refpp{refKitpc}{A.V. Kitaev}{private communication}{1991}
 \refjl{refLukash}{N.A. Lukashevich}{Diff. Eqns.}{3}{395--399}{1967}
\refjl{refMurata}{Y. Murata}{Funkcial. Ekvac.}{28}{1--32}{1985}
\refjl{refNP}{F.W. Nijhoff and V. Papageorgiou}{Phys. Lett.
A}{153}{337--344}{1991}
\refjl{refPNGR}{V. Papageorgiou, F.W. Nijhoff, B. Grammaticos and A.
Ramani}{Phys. Lett. A}{164}{57--64}{1992}
\refjl{refPS}{V. Periwal and D. Shewitz}{Phys. Rev.
Lett.}{64}{1326--1329}{1990}
\refjl{refRGH}{A. Ramani, B. Grammaticos and J. Hietarinta}{Phys. Rev.
Lett.}{67}{1829--1832}{1991}
\refbk{refSchiff}{L.I. Schiff}{Quantum Mechanics}{McGraw-Hill, New
York}{1955}
\refjl{refTGR}{K.M. Tamizhmani,\ B. Grammaticos and A. Ramani}{Lett.
Math.
Phys.}{29}{49--54}{1993}

}

\def\address#1#2{\bigskip{\sc #1}\smallskip
{{\it Email address}: {\tt #2}}\smallskip}

\vskip .5in
\address{Department of Mathematics, University of Exeter, Exeter, EX4
4QE,
U.K.}{clarkson@maths.exeter.ac.uk}
{\it Address from April 1995\/}: Institute of Mathematics \&
Statistics,
University of Kent, Canterbury, CT2 7NF, U.K.
\bigskip
\address{School of Mathematics, University of New South Wales,
Sydney 2052, Australia}{drew@maths.exeter.ac.uk}
{\it Permanent address\/}: Department of
Mathematics,  University of Exeter, Exeter, EX4 4QE, U.K.
\vfill
\date

\bye